# The detection of marine microseismic activity with the CUORE tonne-scale cryogenic experiment

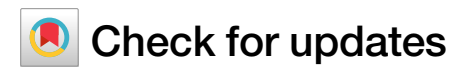

**A list of authors and their affiliations appears at the end of the paper**

Vibrations from experimental setups and the environment are a persistent source of noise for low-temperature calorimeters searching for rare events, including neutrinoless double beta ($0\nu\beta\beta$) decay or dark matter interactions. Such noise can significantly limit experimental sensitivity to the physics case under investigation. Here, we report the detection of marine microseismic vibrations using mK-scale calorimeters. This study employs a multi-device analysis correlating data from CUORE, the leading experiment in the search for $0\nu\beta\beta$ decay with mK-scale calorimeters, and the Copernicus Earth Observation program, revealing the seasonal impact of Mediterranean Sea activity on CUORE's energy thresholds, resolution, and sensitivity over four years. The detection of marine microseisms underscores the need to address faint environmental noise in ultra-sensitive experiments. Understanding how such noise couples to the detector and developing mitigation strategies is essential for next-generation experiments. We demonstrate one such strategy: a noise decorrelation algorithm implemented in CUORE using auxiliary sensors, which reduces vibrational noise and improves detector performance. Enhancing sensitivity to $0\nu\beta\beta$ decay and to rare events with low-energy signatures requires identifying unresolved noise sources, advancing noise reduction methods, and improving vibration suppression systems, all of which inform the design of next-generation rare event experiments.

Low-temperature calorimeters operated at the mK-scale are widely employed for rare physics event searches and for precision measurements. Their broad selection of materials, sizes, and read-out technologies makes them highly suited for a wide range of scientific endeavors, including dark matter and neutrinoless double beta ($0\nu\beta\beta$) decay searches, supernova neutrinos detection through coherent elastic neutrino-nucleus scattering, direct measurements of neutrino masses, and $\beta$ decay shape studies[1,2]. $0\nu\beta\beta$ decay[3] is a hypothetical rare nuclear process whose discovery would provide insights into physics beyond the Standard Model and into some of the most outstanding mysteries of our Universe. It would establish the nature of neutrinos as Majorana particles, meaning that, uniquely among all known fundamental fermions, they would be indistinguishable from their own antiparticles[4]. Moreover, it would provide the first evidence of a process violating lepton number conservation. Such processes have the potential to explain the matter-antimatter asymmetry in the Universe via baryogenesis[5].

An energy deposition in a low-temperature calorimeter generates phonon excitations, resulting in a measurable increase of the detector temperature, which is then converted into an electric signal by means of a thermal sensor[6]. The CUORE experiment[7], located at Laboratori Nazionali del Gran Sasso (LNGS) of the Istituto Nazionale di Fisica Nucleare (INFN), in Italy, searches for $0\nu\beta\beta$ decay in $^{130}$Te[8], by deploying a tonne-scale array of low-temperature calorimeters operated at the mK-scale. CUORE consists of an array of 988 $^{nat}TeO_2$ crystals ($5 \times 5 \times 5\ cm^3$, 750 g), read out by high-impedance germanium Neutron Transmutation Doped (Ge-NTD) thermistors and operated as low-temperature calorimeters at ≃10 mK (see Fig. 1). Such technology allows one to achieve cutting-edge energy resolutions (≃0.3% FWHM at ~2.5 MeV, the typical scale of $\beta\beta$ decay Q-value) and radio-purity levels, as well as to deploy active masses up to the tonne-scale. CUORE has been in operation since 2017, reaching an outstanding duty cycle (>90% since 2019) for a tonne-scale experiment operating at mK temperature, and acquiring >2 tonne·yr exposure of $TeO_2$, the highest ever achieved for $^{130}$Te[9].

Several noise sources, including extrinsic vibrations and electronic interference, can affect the performance of low-temperature calorimeters operated at the mK-scale. In fact, intermittent power deposition (e.g., due to vibrations) of ~1–10 fW can result in transient noise signals in the calorimeters. If this excess noise is not time-invariant, it can worsen the energy

e-mail: cuore-spokesperson@lngs.infn.it

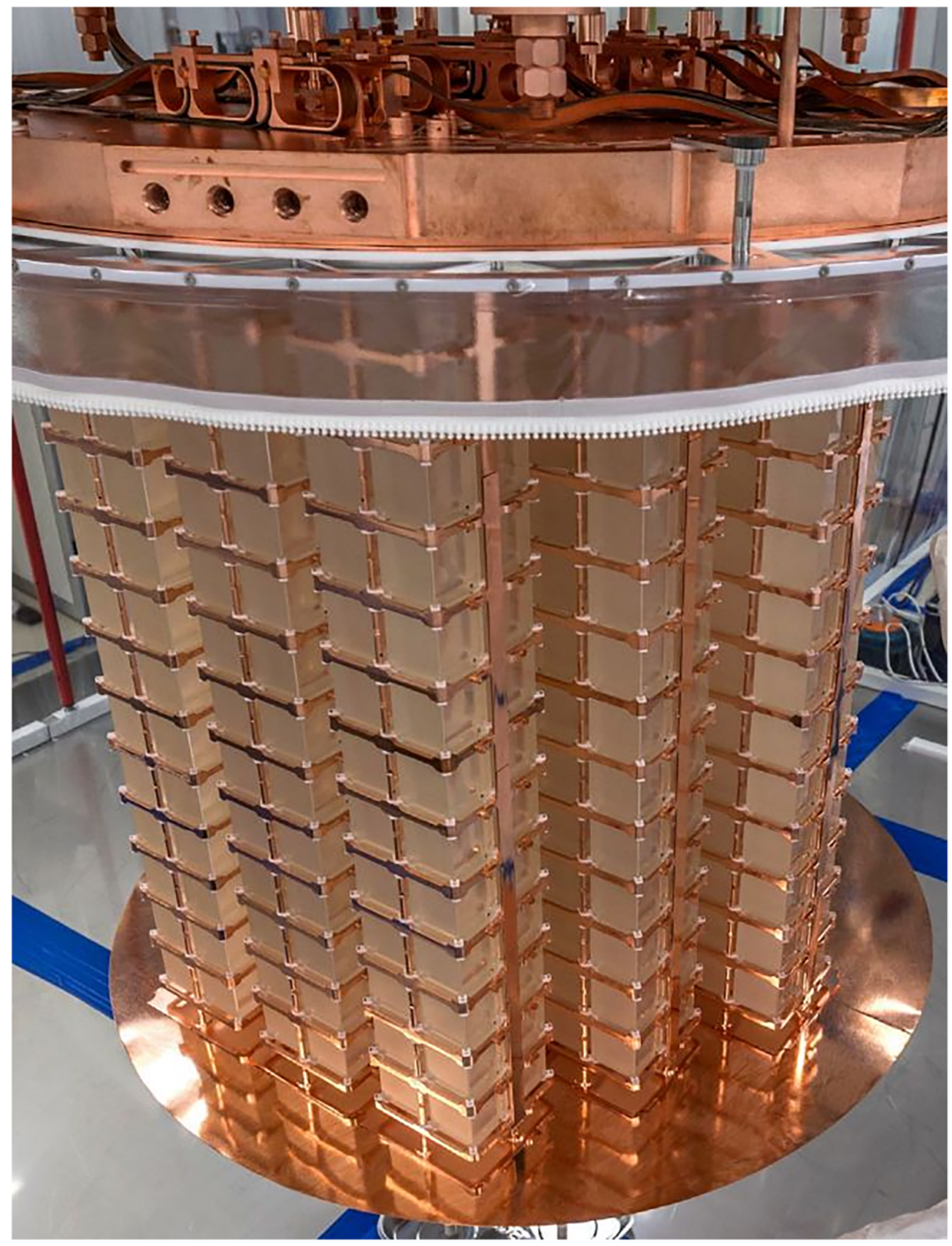

**Fig. 1 | The CUORE experiment.** The active volume of the CUORE experiment, consisting of 988 low-temperature calorimeters organized in 19 towers of 13 floors, each of them hosting 4 detectors.

resolution of the detectors, as the matched filtering technique only produces the optimal energy resolution in the case of a linear, time-invariant system[10]. In CUORE, the high-impedance calorimeters are highly sensitive to noise perturbations with frequencies up to several hundred Hz, and maximally sensitive in the sub-Hz regime, where the thermal signal band lies. Since each detector is read out via gold wires bonded to copper strips deposited on PEN bands running along the detector frame, oscillations transmitted through the read-out wires can induce an excess of noise in the detectors via capacitive pick-up.

Vibrational noise from the cryogenic system, anthropogenic activities, and the environment (e.g., microseismic activity, earthquakes, and sea swell motion) plays a crucial role in defining the performance of low-temperature calorimeters. Therefore, such detectors must implement various strategies to mitigate vibrational noise. The CUORE detector array is hosted within a custom $^3$He–$^4$He dilution cryostat and is suspended by means of a mechanical decoupling system. This system is designed to mitigate the impact of vibrations from both the external environment and the cryogenic infrastructure required to operate the detectors at ≃10 mK[7,11]. Additionally, a noise cancellation technique has been developed to suppress vibrational noise at harmonics of 1.4 Hz induced by the operation of the pulse tube (PT) cryocoolers of the CUORE cryostat[12]. This ensures enhanced temperature stability ($\frac{\Delta T}{T}$ < 1% at 10 mK) and the mitigation of low-frequency vibrations induced in the detectors.

We present the detection of marine microseisms through low-temperature calorimeters. By correlating data from the Copernicus Earth Observation space program, from seismometers installed at LNGS, and from CUORE low-temperature calorimeters, we identify the contribution to the CUORE energy resolution from marine microseismic vibrations. We assess a correlation between the seasonal modulation of the Mediterranean Sea activity and the induced modulation of the energy resolution of CUORE detectors, while also estimating the corresponding impact on the experimental sensitivity to the search for $0\nu\beta\beta$ decay.

We then present a denoising algorithm, which combines data from low-temperature calorimeters and auxiliary devices, including accelerometers and a seismometer, for vibrational noise reduction. The implementation of the denoising algorithm, in synergy with various passive and active noise reduction techniques, allows the CUORE detectors to reach previously unattained levels of noise.

## Results

### Detecting marine microseisms with CUORE

The noise reduction achieved via passive and active noise cancellation techniques makes CUORE highly sensitive to subtle sources of environmental vibrations[13], which would otherwise be subdominant. The motion of sea swells is known to be a faint and everlasting source of microseismic vibrations in the sub-Hz domain, capable of propagating from seas and oceans to the mainland[14–17]. However, its impact on the performance of cryogenic experiments dedicated to rare event searches had yet to be investigated. Located approximately 50 km and 150 km from the Adriatic and Tyrrhenian coastlines of the Italian Peninsula, respectively, LNGS is affected by the subtle microseismic activity induced by the Mediterranean Sea.

We report the outcomes of a multi-device correlation analysis involving:

- Copernicus Marine Environment Monitoring Service (CMEMS), namely the marine component of Copernicus, the European Union space programme for Earth Observation[18,19], providing state-of-the-art marine data at global and regional scale;
- seismometers installed near the CUORE infrastructure and ≃130 m away, at the opposite side of the LNGS underground facility;
- CUORE low-temperature calorimeters.

With respect to the preliminary investigation presented in ref. 20, the analysis framework presented here accounts for the entire CUORE array of low-temperature calorimeters, and for four years of steady data acquisition. We show that variations in the microseismic activity of the Mediterranean Sea induce a sizeable effect on the performance of the whole CUORE array in terms of the detectors' low-energy thresholds and energy resolution. We highlight such correlations both on the time scale of the outbreak of Mediterranean storms (≃1–2 weeks) and over years, thus assessing seasonal patterns.

Figure 2 (b) shows the comparison between the time profile of the summed sea wave amplitude (VHM0) in the Adriatic and Tyrrhenian Seas and the time profile of the seismic activity at the LNGS underground facility, both measured during a storm outbreak (see picture from Copernicus in Fig. 2(a)). The high correlation between them highlights that the (micro) seismic activity at LNGS increases during the outbreak of storms in the Mediterranean region. The characteristic sub-Hz frequencies of microseismic vibrations fall within the signal band of the CUORE detectors, therefore contributing to their noise[20]. Figure 2(c) shows the linear correlation between the Mediterranean sea activity $I_S$ and the baseline resolution $\mathrm{FWHM}_{\mathrm{baseline}}$ of CUORE detectors. We define the sea activity as the time integral of the summed sea wave amplitude in the Adriatic and Tyrrhenian Seas, while the baseline resolution of a low-temperature calorimeter is defined as the energy resolution at zero released energy. Since it quantifies the contribution to the total energy resolution due solely to noise fluctuations, the baseline resolution is a measure of the overall noise level of the system.

The correlation chain between marine, seismic, and calorimetric data implies that the excess of microseismic activity during storm outbreaks, with typical durations of ≃1–2 weeks, can induce a sizable worsening in the energy resolution of detectors operating at the mK-scale (up to ≃40% during the storms reported in Fig. 2). We highlight the role of seismic data acquired underground at LNGS as the bridge allowing us to directly correlate marine data from CMEMS and calorimetric data from CUORE, as shown in Fig. 2.

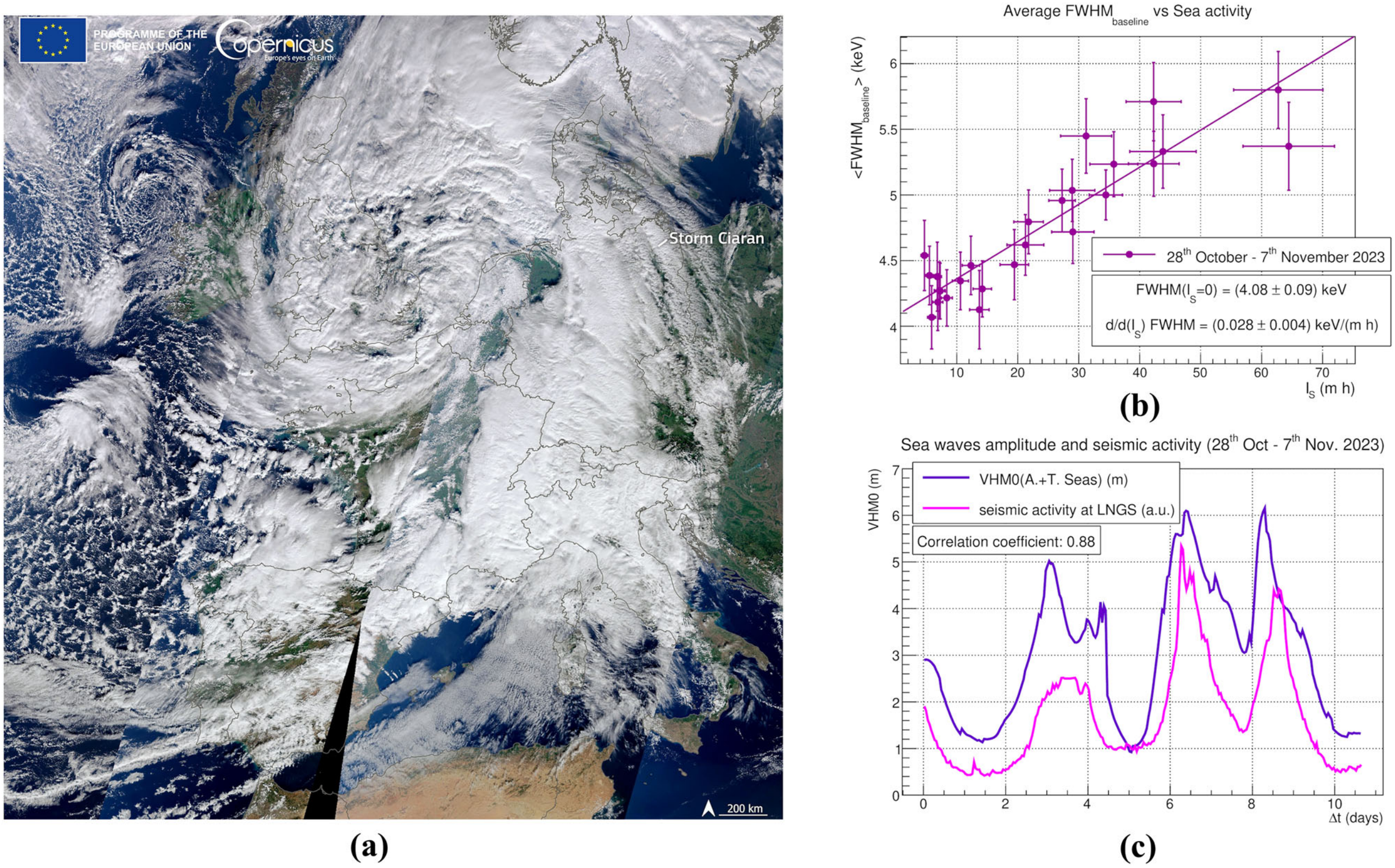

**Fig. 2 | Correlations between marine, seismic, and calorimetric data. a** Picture of a storm engulfing Western Europe on 2nd November 2023 (courtesy of Copernicus Earth observation space program[19]). **b** Time evolutions of the summed sea wave amplitude (VHM0) in the Adriatic and Tyrrhenian Seas, and of the seismic activity at LNGS, during the storm outbreak. The Pearson correlation coefficient between the two time profiles is reported in the legend. **c** Linear correlation between the Mediterranean sea activity $I_S$ (Eq. (3)) and the average CUORE baseline resolution $\mathrm{FWHM}_{\mathrm{baseline}}$ during the storm outbreak. Each data point accounts for ≃12 h of Copernicus and CUORE data. The systematic uncertainty on $I_S$ is estimated as the sea activity during the first and last hour of each ≃12 h-long time period, accounting for the time mismatch between the beginning/end of each set of Copernicus data, available with a time resolution of 1 h, and of CUORE data. The average baseline resolution is evaluated on detectors hosted in the upper five floors of CUORE towers, as representative of the most sensitive detectors to variations in microseismic noise. The corresponding uncertainty is estimated as the mean value uncertainty. The solid line represents the linear fit to the data, while the best-fit parameters are reported in the legend.

## Seasonal modulation of the CUORE performance

The stable data-taking of CUORE since 2019 offers the unique opportunity to unveil the interplay between low-temperature calorimeters and environmental phenomena over many years. It is well assessed that the Mediterranean Sea activity modulates seasonally due to more frequent and intense storms occurring during winter compared to summer[21]. Consequently, the intensity of the induced microseismic noise is also season-dependent. By extending our investigation over many years, we assess a seasonal modulation of the detectors' low-energy threshold and energy resolution at the 2615 keV $^{208}$Tl $\gamma$-ray peak, being crucial parameters defining, respectively, the experimental sensitivity to rare events with low-energy experimental signatures and to $0\nu\beta\beta$ decay.

The baseline resolution of a low-temperature calorimeter directly determines its low-energy threshold, namely the minimum energy deposition that can be discriminated from noise fluctuations of the baseline. After evaluating the threshold of each CUORE low-temperature calorimeter, we define the low-energy mass exposure as the total active mass of detectors achieving a threshold lower than 10 keV. Such a threshold is a base requirement to search for rare events with experimental signatures in the low-energy part of the spectrum[22,23]. Figure 3 shows the comparison between the Mediterranean Sea wave amplitude and the low-energy mass exposure over four years, from January 2019 to April 2023. The sea wave amplitude is averaged over every 2 months; this time resolution allows us to scan the seasonal variations of the Mediterranean Sea while washing out transient storms, so that average seasonal patterns can emerge. The low-energy mass exposure of the entire CUORE detector array is evaluated over similar periods after applying the analysis procedure described in ref.[24].

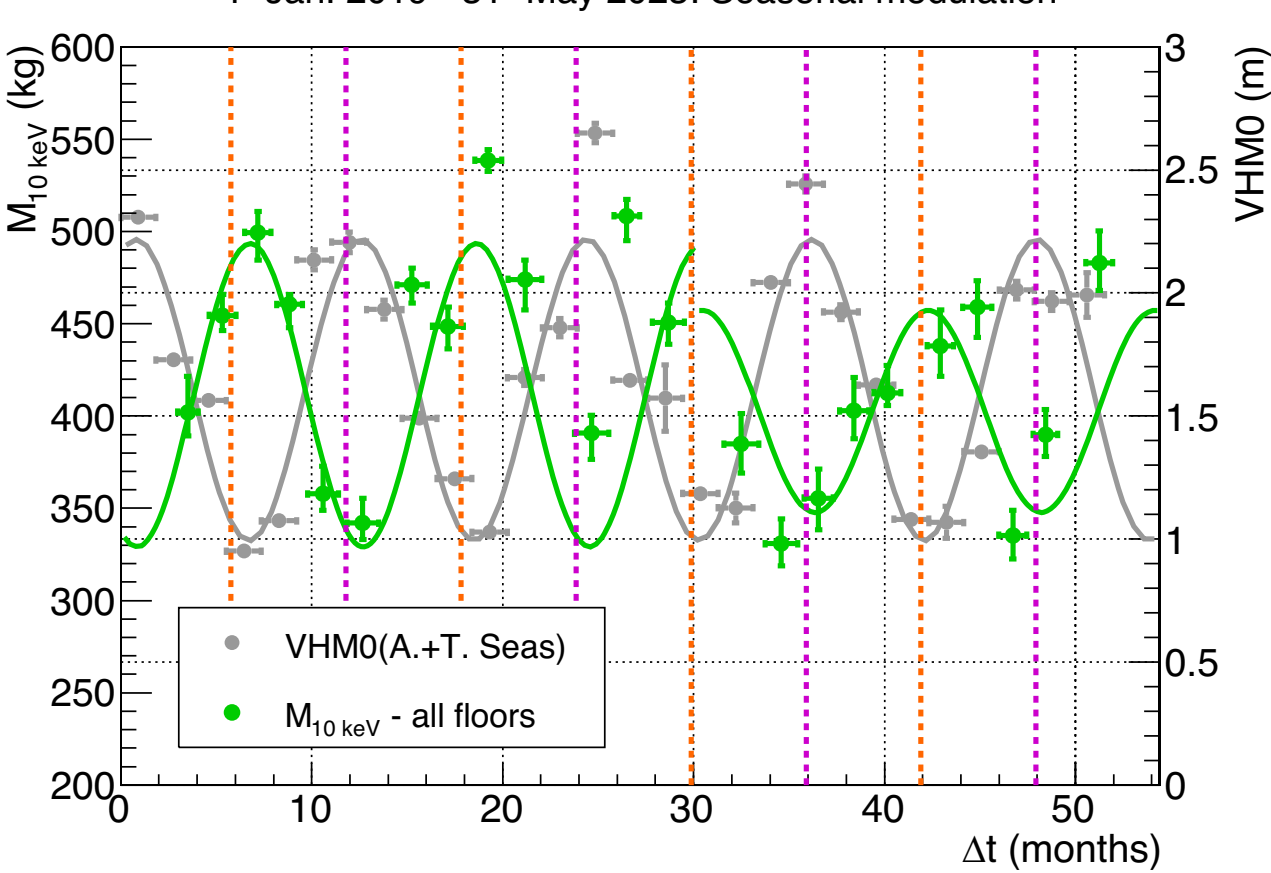

**Fig. 3 | Modulation of CUORE low-energy exposure driven by the Mediterranean Sea activity.** Seasonal modulations of the Mediterranean Sea wave amplitude (VHM0) and of the mass exposure below 10 keV ($M_{10\ \mathrm{keV}}$) of the entire CUORE detector array. The time axis starts on January 1st 2019. The error bars along the time axis represent the time intervals over which VHM0 and the CUORE exposure are evaluated. Continuous lines represent the outcome of a simultaneous fit procedure. Vertical dashed lines refer to summer (orange) and winter (purple) solstices. In July 2021 (30th month), the base temperature of CUORE was changed from 11.8 mK to 15.0 mK.

**Table 1 | Seasonal modulation parameters for: Mediterranean Sea wave amplitude (VHM0); CUORE mass exposure below 10 keV ($M_{10\ keV}$); CUORE FWHM energy resolution at $^{208}$Tl(2615 keV) $\gamma$ peak. The reduced $\chi^2$ is also reported as a metric of the goodness of fit**

| | VHM0 | $M_{10keV}$ @ 11.8 mK | $M_{10keV}$ @ 15.0 mK | FWHM($^{208}$Tl) |
|---|---|---|---|---|
| $A$ | (0.63 ± 0.05) m | (82.2 ± 10.6) kg | (54.8 ± 9.6) kg | (0.61 ± 0.09) keV |
| $T$ | (11.8 ± 0.2) months | | | |
| ϕ | 1.2 ± 0.2 | $(1.2 + \pi)$ ± 0.2 | | 1.2 ± 0.2 |
| $C$ | (1.65 ± 0.04) m | (411.3 ± 6.8) kg | (402.5 ± 6.8) kg | (7.34 ± 0.10) keV |
| $\chi^2_{red}$ | 1.39 | | | |

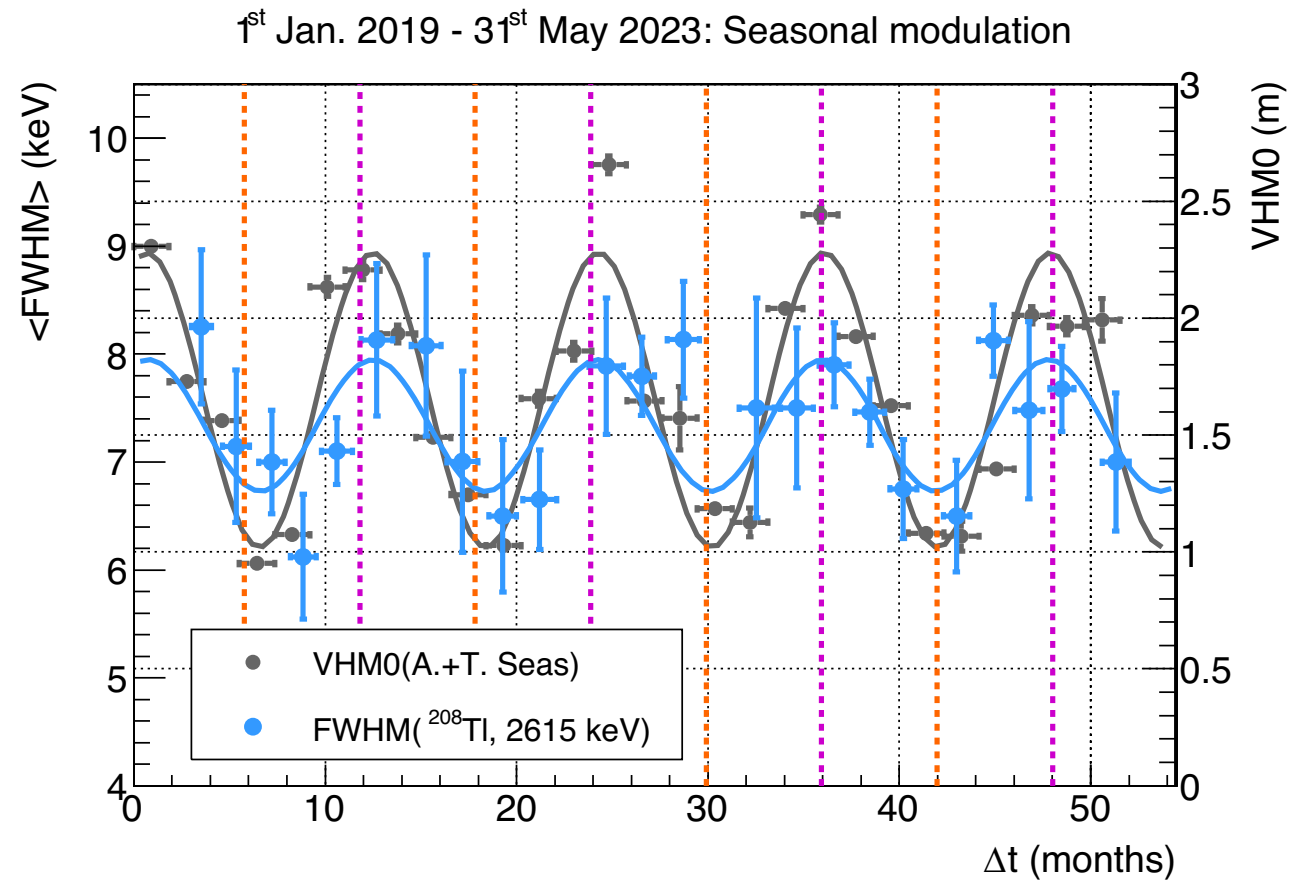

**Fig. 4 | Modulation of CUORE energy resolution driven by the Mediterranean Sea activity.** Seasonal modulations of the Mediterranean Sea wave amplitude (VHM0) and of the energy resolution of the entire CUORE detector array at $^{208}$Tl $\gamma$-ray peak. The time axis starts on January 1st, 2019. The error bars along the time axis represent the time intervals over which VHM0 and the CUORE energy resolution are averaged. Continuous lines represent the outcome of a simultaneous fit procedure. Vertical dashed lines refer to summer (orange) and winter (purple) solstices. In July 2021 (30th month), the base temperature of CUORE was changed from 11.8 mK to 15.0 mK. The fit procedure does not differentiate between the energy resolutions measured before and after the temperature change, given that the energy resolution of CUORE is not dominated by thermal effects.

CUORE's low-energy mass exposure shows an annual modulation with a phase opposite to that of the Mediterranean Sea wave amplitude variations (see Table 1). The higher Mediterranean Sea activity in the winter months worsens the CUORE detectors' thresholds, resulting in a lower mass exposure below 10 keV. The number of detectors achieving a threshold lower than 10 keV drops from ≃76% of the total during the summer minimum microseismic activity to ≃48% during the winter maximum activity. We estimate a maximum summer-to-winter variation in the low-energy mass exposure of ≃165 kg when the CUORE detectors are operated at 11.8 mK, and of ≃110 kg when they are operated at 15.0 mK, corresponding to a 40.0% and 27.2% variation in the average mass exposure.

Figure 4 shows the comparison between the Mediterranean Sea wave amplitude and the energy resolution at the 2615 keV $^{208}$Tl $\gamma$-ray peak from January 2019 to April 2023. The $^{208}$Tl FWHM energy resolution of the entire CUORE detector array is evaluated after applying the analysis procedure described in ref.[25]. CUORE's $^{208}$Tl energy resolution exhibits a seasonal modulation characterized by an in-phase yearly periodicity consistent with the variations of the Mediterranean Sea wave amplitude, with minima observed in summer and maxima in winter. Indeed, the higher activity of the Mediterranean Sea in winter compared to summer worsens the detector's energy resolution. Near the region of interest for $0\nu\beta\beta$ decay searches in $^{130}$Te ($Q_{\beta\beta}$=2527.5 keV), the variation in CUORE's energy resolution over the seasons exceeds 1 keV. Finally, the average energy resolution aligns with the value extrapolated from the CUORE cumulative time-integrated analysis[9].

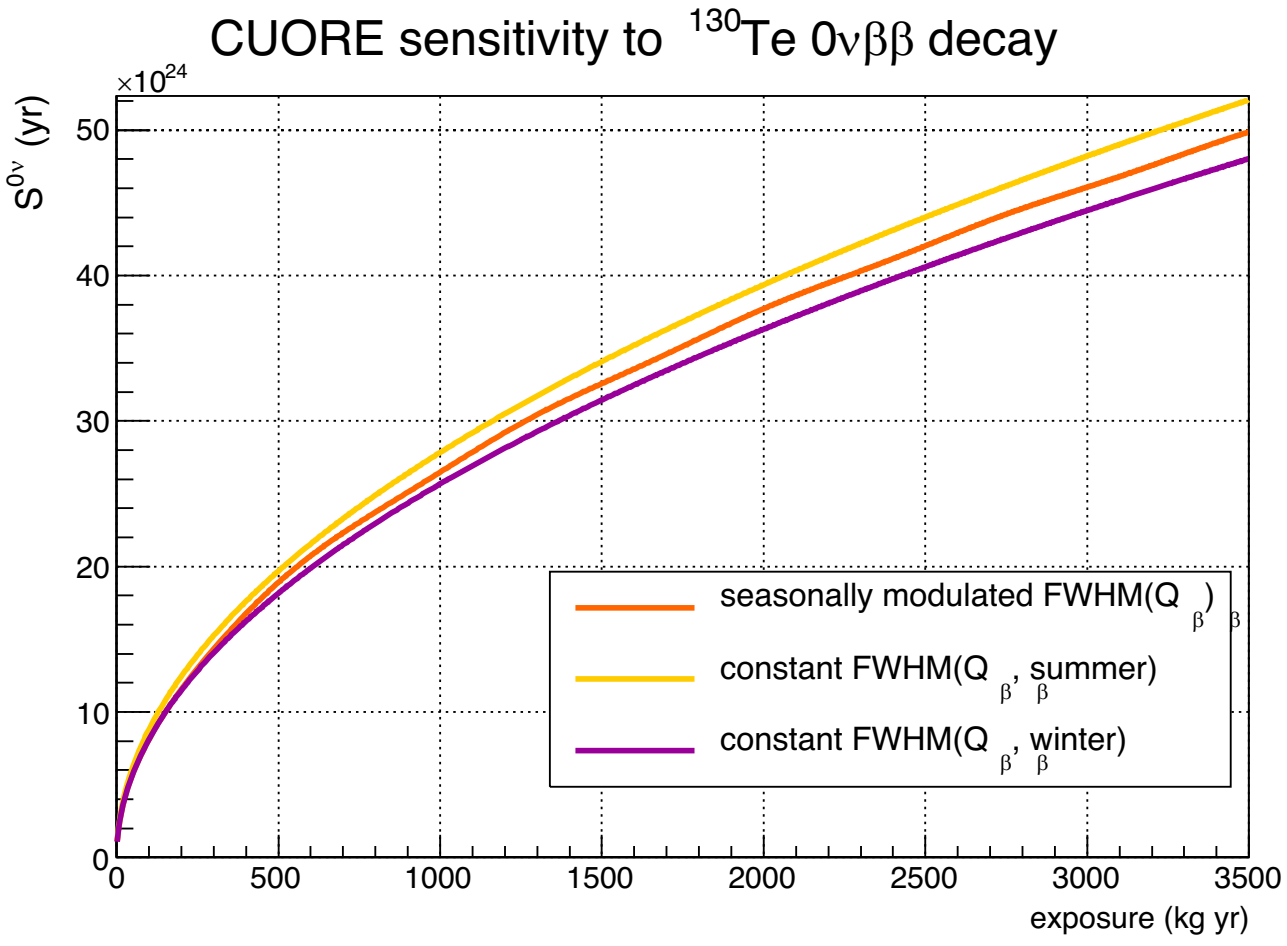

**Fig. 5 | Sensitivity of CUORE experiment to neutrinoless double beta decay.** Cumulative sensitivity of CUORE to $0\nu\beta\beta$ decay half-life assuming different energy resolutions, namely the annual average, the winter worst, and the summer best resolutions.

The worsening of CUORE's energy resolution due to microseismic activity translates into a reduction of its sensitivity to $0\nu\beta\beta$ decay for a given experimental exposure. In order to estimate the loss in $0\nu\beta\beta$ decay sensitivity due to microseismic noise, we compare the actual CUORE sensitivity $S^{0\nu}$, accounting for a seasonally modulated energy resolution, with the sensitivity $S^{0\nu}_S$ that would be achieved if CUORE's energy resolution were constant and equal to the one measured at the summer minimum sea activity (see Fig. 5). The latter case is the closest data-driven approximation of the ideal scenario of CUORE being fully decoupled from microseismic vibrations.

The estimated sensitivity loss between these two scenarios is:

$$\frac{S^{0\nu}_S - S^{0\nu}}{S^{0\nu}} \gtrsim 4.3\% \quad (1)$$

This result underscores the critical importance of mitigating microseismic effects to unlock CUORE's full potential for year-round sensitivity to $0\nu\beta\beta$ decay and potential other rare events.

## Denoising technique in CUORE

CUORE is equipped with various auxiliary devices (seismometers, accelerometers, and microphones) positioned at several external points of the cryogenic and mechanical suspension infrastructure. These devices enable monitoring of the response of the experimental infrastructure to vibrations. The role of seismometers is crucial thanks to their sensitivity to vibrations in the sub-Hz domain, including marine microseismic vibrations, which lie within the CUORE signal band.

Leveraging the correlation between the noise measured by the auxiliary instruments and the response of the low-temperature calorimeters, we recently developed a denoising algorithm that can be implemented during post-processing of the data[26]. The denoising procedure enhances the noise stability over time by constructing the denoising correlation function on a daily basis, and lowers the trigger thresholds for enhanced sensitivity to experimental signatures in the low-energy portion of the energy spectrum[9]. This approach has been previously tested on ≃24 h of data from a single persistently noisy CUORE calorimeter, and it proved effective in significantly reducing the vibrational noise[26]. However, its efficacy had yet to be demonstrated on the entire CUORE detector array and over longer time scales.

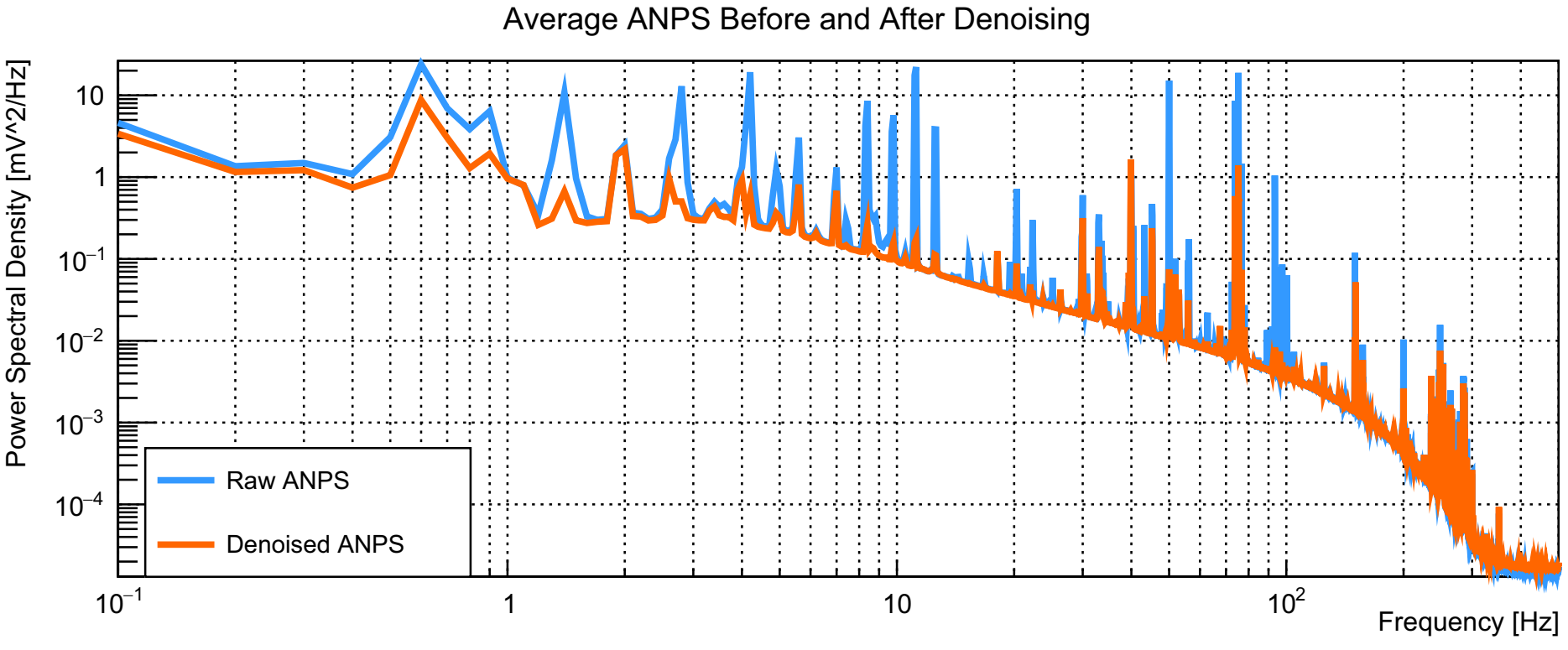

**Fig. 6 | Impact of the denoising algorithm on the whole CUORE detector.** Average noise power spectrum (ANPS) averaged over all CUORE detectors before (blue) and after (orange) applying the denoising algorithm. The denoising reduces the total noise power, primarily mitigating harmonics of 1.4 Hz associated with PT noise and sub-Hz peaks correlated with sea wave activity.

Here we show the effect of the denoising algorithm on the full CUORE detector array over a reference two-month period, during which the vibrational noise in the immediate surroundings of the experimental infrastructure was measured by one seismometer, two accelerometers, and two microphones. We selected this period because the seismometer was installed at a later time with respect to the other auxiliary devices. The seismometer has the highest sensitivity to the sub-Hz vibrational noise correlated with the sea activity, given its sensitivity band (~0.1–50 Hz) and sensitivity (≃400 V/(m/s)). The accelerometers are characterized by the same frequency band (~0.1–50 Hz), but a lower sensitivity (≃1 V/(m/s$^2$)), while the microphones feature a sensitivity band roughly equivalent to the range of human hearing (~20 Hz–20 kHz).

We perform two parallel analyses to evaluate the performance of the denoising technique. In the first analysis, the data are processed directly in their raw form, without implementing the denoising algorithm. In the second analysis, the denoising algorithm is applied as the initial step in the data processing pipeline.

Figure 6 shows the average noise power spectrum (ANPS) averaged over all detectors before and after the denoising is applied. The algorithm reduces the total noise power by 74%, with the vast majority of the reduction occurring at harmonics of 1.4 Hz, where the PT noise dominates. For instance, the 1.4 Hz noise peak is reduced by more than 12 decibels. The noise power in the frequency interval [0.1, 1] Hz is reduced by 56%. The highest relative noise reduction in this region occurs at the 0.6 Hz and 0.9 Hz peaks, which are correlated with the sea wave activity[20].

CUORE applies an optimum filter (OF) technique[10,25] to reconstruct the amplitude of the thermal signals; this approach is effective in filtering away time-independent noise components. In contrast, the denoising algorithm suppresses both time-independent and transient noise by predicting the detector response to mechanical vibrations using the auxiliary device signals. We quantify the impact of the denoising technique on the detector resolution by evaluating the expected baseline amplitude resolution after applying the OF:

$$\sigma_A = \left[ T \sum_{i=1}^{n-1} \frac{|s[f_i]|^2}{N[f_i]} \right]^{-1/2} \tag{2}$$

where $\sigma_A$ is the expected amplitude resolution (std. dev.) of the detector after applying the OF, $s[f]$ is the average pulse (i.e., the signal template) with amplitude 1, $N[f]$ is the ANPS, and $T$ is the duration of the average pulse and the noise window. The sum is performed over all $n$ frequency bins except the DC component. Note that the amplitude resolution is proportional to the baseline energy resolution of the given detector, but the constant of proportionality is detector-dependent.

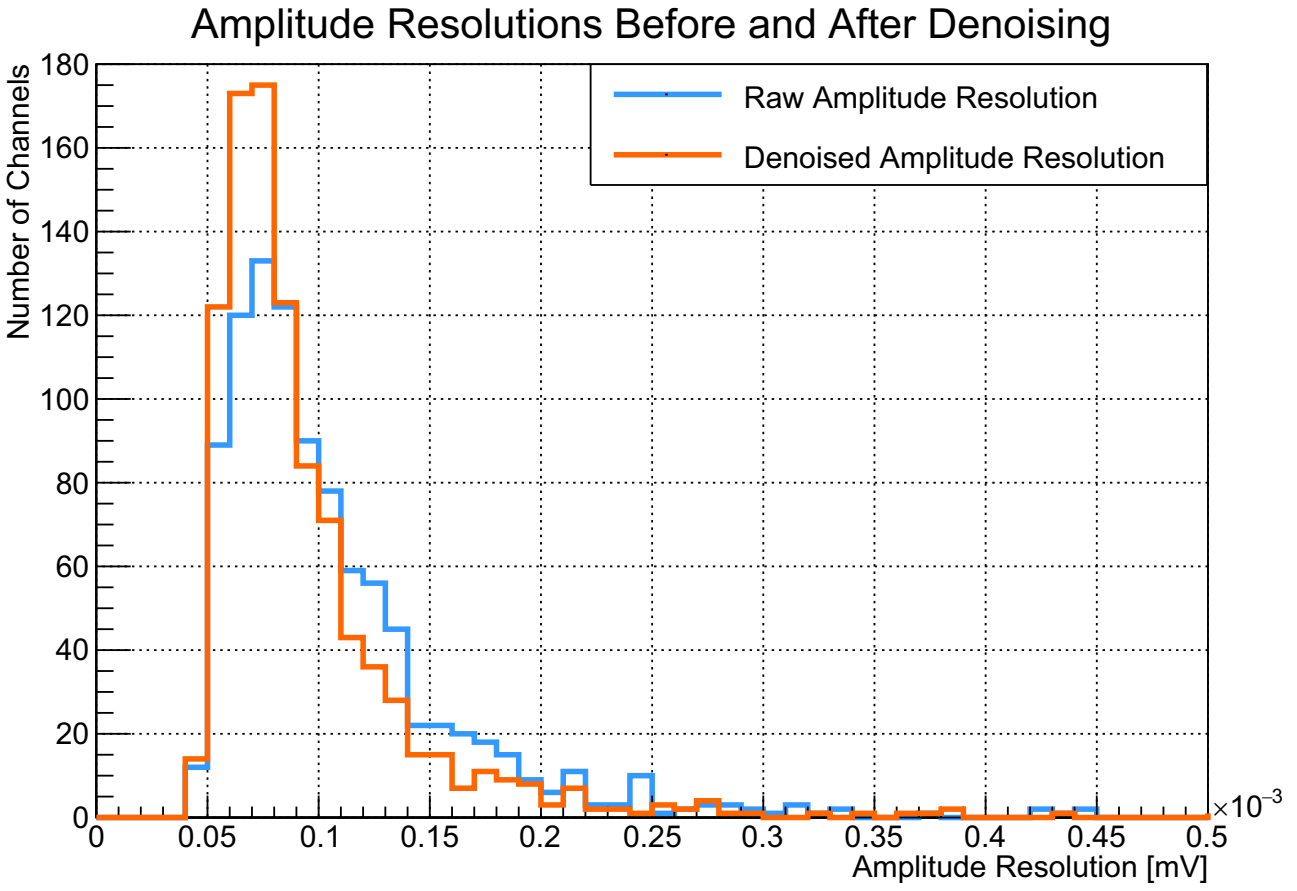

**Fig. 7 | Amplitude resolutions before and after denoising.** Distribution of amplitude resolutions $\sigma_A$ for CUORE detectors before denoising (blue) and after denoising (orange). The average amplitude resolution improves by 11.9%, and the median resolution improves by 11.8%.

Figure 7 shows the distribution of expected amplitude resolutions for each detector in CUORE before and after applying the denoising. The denoising technique improves the baseline resolution in 94% of the analyzed detectors for a mean improvement of 11.9%, respectively. Since the baseline resolution is directly proportional to the detector's energy threshold, the denoising technique lowers the threshold of 94% of channels as well. The benefits of such improvements include the possibility of reducing the background at $Q_{\beta\beta}$, thanks to the improvement in tagging time-coincident events, and of increasing the sensitivity to physics processes with low-energy signatures.

## Discussion

We have demonstrated that the CUORE experiment is sensitive to the sub-Hz microseismic vibrations induced by the everlasting activity of the Mediterranean Sea. We assess and quantify the impact of microseismic noise on crucial parameters for low-temperature calorimeters operated at the mK scale, namely their energy resolutions and thresholds. The seasonal variation of the Mediterranean Sea activity reflects a seasonal modulation of the CUORE energy resolution, which affects the experimental sensitivity to $0\nu\beta\beta$ decay. Such a correlation between subtle environmental phenomena and CUORE performance represents proof of the cutting-edge performance achievable with low-temperature calorimeters.

The assessment of the interplay between microseismic phenomena and low-temperature calorimeters highlights the need for improvements in noise reduction algorithms. In this regard, we demonstrated the benefits of a denoising algorithm that mitigates the residual vibrational noise affecting the CUORE detectors, improving the baseline stability and resolution, thereby enhancing the sensitivity to rare processes with experimental signatures in the low-energy range of the spectrum. Moreover, this denoising technique can potentially be implemented in any experiment that requires the mitigation of vibrational noise.

The identification of previously unresolved sources of noise, the improvement of noise reduction algorithms, and further developments in vibration suppression systems could guide the design of next-generation experiments with enhanced sensitivity to rare events, including CUPID, a next-generation tonne-scale experiment for $0\nu\beta\beta$ decay searches with mK-calorimeters[27].

## Methods

### Copernicus Marine Environment Monitoring Service

CMEMS data are based on satellites and in-situ data and on numerical models[28,29] of the marine and atmospheric environments. The spectral significant wave height (VHM0) is the average of the highest one-third of wave heights, historically defined to correspond to the intuitive visual measure of the wave height by seafarers. CMEMS data are available with a time resolution of 1 h and a spatial resolution of $\simeq 4.6$ km. In this work, we focus on two domains in the Adriatic ([41.5, 46.0]°N × [12.6, 19.6]°E) and Tyrrhenian ([38.0, 42.5]°N × [9.7, 14.2]°E) Seas, over which we evaluate the hourly average VHM0 value. The definition of different sea domains does not affect the estimation of the average VHM0[20].

In this analysis, we define the sea activity $I_S$ as:

$$I_S = \int_{t_i}^{t_f} \left[\mathrm{VHM0}_A(t) + \mathrm{VHM0}_T(t)\right] dt \tag{3}$$

where the labels $A$ and $T$ refer respectively to the Adriatic and Tyrrhenian Seas, and $\left[t_i, t_f\right]$ are time periods of $\simeq 12$ h.

### Interpolation procedure for seasonal modulations

The solid lines in Figs. 3 and 4 represent the outcome of a combined fit procedure of CMEMS and CUORE data. The fit procedure is performed through the BAT (Bayesian Analysis Toolkit) software[30]. We define a Gaussian likelihood, which approximates the statistics of the two variables, whose mean value $\mu$(t) is defined by a sinusoidal function, describing their seasonal variations over time:

$$\mu(t) = A \sin\left(\frac{2\pi}{T} t + \phi\right) + C \tag{4}$$

The oscillation period $T$ is defined as a common fit parameter for both CMEMS and CUORE data, while the oscillation amplitude $A$ and the average value $C$ are independent degrees of freedom. The phase $\phi$ is also defined as a common fit parameter between CMEMS and CUORE data. Consequently, we introduced a fixed phase shift of $\pi$ to the modulation of the low-energy mass exposure, in order to account for its counter-phase pattern with respect to the modulation of the sea wave amplitude (Fig. 3). The outcomes of the maximum-likelihood fit procedures are reported in Tab. 1, where the uncertainties on fit parameters are defined as the $1\sigma$ confidence level of the marginalized posteriors.

The energy resolution of CUORE at $Q_{\beta\beta}$ is estimated using a scaling function describing the detector resolution over the whole energy range of CUORE, spanning from keV to MeV scale[9,25]. By scaling the energy resolution at $^{208}$Tl peak during the different seasons, the corresponding resolutions at $Q_{\beta\beta}$ can be evaluated. The relative seasonal variation of the energy resolution at $Q_{\beta\beta}$ is found to be:

$$\frac{\mathrm{FWHM}_W(Q_{\beta\beta}) - \mathrm{FWHM}_S(Q_{\beta\beta})}{\mathrm{FWHM}_S(Q_{\beta\beta})} \simeq 20\% \tag{5}$$

where $W$ and $S$ index winter and summer, respectively.

The energy resolution is a crucial parameter defining the experimental sensitivity to $0\nu\beta\beta$ decay, whose signature is a mono-energetic peak in the summed energy spectrum of the two emitted electrons at the decay Q-value ($Q_{\beta\beta}$). The sensitivity $S^{0\nu}$ of a background-limited experiment is described by the figure of merit[31]:

$$S^{0\nu} \propto \sqrt{\frac{M \cdot T}{\langle\,\mathrm{FWHM(T)}\,\rangle \cdot B}} \tag{6}$$

where $M$ is the active mass of the $0\nu\beta\beta$ candidate isotope, $T$ is the measurement time, $\langle$FWHM(T)$\rangle$ is the time-averaged energy resolution at $Q_{\beta\beta}$ and $B$ is the background index around it. The time-averaged energy resolution is defined as:

$$\langle\,\mathrm{FWHM(T)}\,\rangle = \frac{1}{T}\int_0^T \mathrm{FWHM}(t)\,\mathrm{d}\,t \tag{7}$$

where FWHM($t$) is the seasonally-modulated energy resolution in Fig. 4.

### The denoising technique

We apply the denoising algorithm independently on each CUORE detector, exploiting different auxiliary devices as inputs. In particular, for the analysis reported in this paper, we employed a triaxially mounted set of three accelerometers, one seismometer, and two microphones. The frequency range of the seismometer spans from < 0.1 Hz to ~100 Hz, that of the accelerometers spans from 0.1 Hz to 200 Hz, and that of the microphones spans from 20 Hz to 20 kHz. Therefore, the array of auxiliary devices covers the entire frequency range of physics signals in the CUORE calorimeters. In addition to the auxiliary device signals, we also consider their squares to account for non-linear effects expected in the noise.

During physics runs, the trigger rate is sufficiently low to identify multiple 50-s time windows during which no pulses are present (noise events), utilized for building the denoising transfer function. The transfer functions are built in the frequency domain using the following expressions:

$$G_{yy}[f] = \frac{2}{T}\langle Y^*[f]Y[f]\rangle \tag{8}$$

$$G_{x_i y}[f] = \frac{2}{T}\langle X_i^*[f]Y[f]\rangle \tag{9}$$

$$G_{x_i x_j}[f] = \frac{2}{T}\langle X_i^*[f]X_j[f]\rangle \tag{10}$$

Here, $T$ is the 50-s window length, $X_i$ is the Fourier transform of the $i$-th input signal, and $Y$ is the Fourier transform of the detector. The expectation values are taken over the sets of noise events. At each frequency, $G_{yy}$ is the value of the ANPS of the detector signal, $G_{x_i y}$ is a vector of cross-spectral densities of the output with each input, and $G_{x_i x_j}$ is a matrix of the cross-spectral densities of the inputs. The on-diagonal terms $G_{x_i x_i}$ comprise the ANPS of the input signals. After averaging over all noise events, the transfer functions from the inputs to the detector are given by ref. 26:

$$H_{x_i y} = G_{x_i x_j}^{-1} G_{x_j y} \tag{11}$$

These transfer functions are then applied to the input devices to produce the predicted detector noise $Y_p$. Subtracting $Y_p$ from $Y$ gives the denoised detector signal. A schematic of the algorithm is shown in Fig. 8.

The data acquired during calibration periods exhibit an event rate 10 times higher than that of the physics runs, resulting in an insufficient number of noise events to construct reliable transfer functions. These data are denoised by averaging the transfer functions from multiple temporally proximate physics runs.

The denoising technique reduces the total noise power in 99.8% of the detectors. The average relative noise power across all detectors after

**Fig. 8 | A schematic of the denoising algorithm.** **a** The transfer functions from input devices to a CUORE calorimeter are constructed from many noise events by means of Eqs. (8) through (11). **b** Once the transfer functions are constructed, they are convolved with the input device signals. The results are summed to produce a predicted calorimeter noise waveform, which is then subtracted from the original waveform. The difference is the denoised waveform.

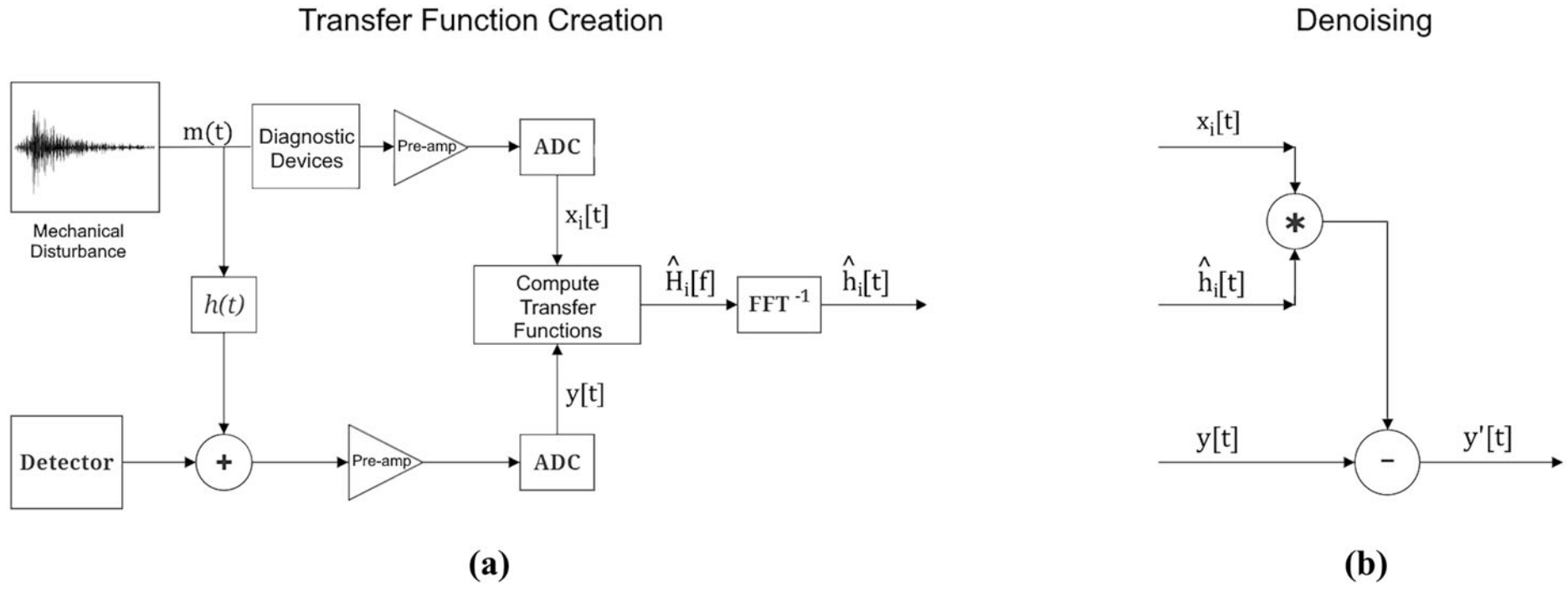

denoising is evaluated as:

$$\frac{1}{n_{det}}\sum_{i=1}^{n_{det}}\frac{\sum_f N_i'[f]\Delta f}{\sum_f N_i[f]\Delta f} = 0.262 \tag{12}$$

where $n_{det}$ is the number of detectors for which an ANPS can be constructed, $N_i'[f]$ is the denoised ANPS of the $i$-th detector, $N_i[f]$ is the original ANPS of the $i$-th detector, and $\Delta f = 0.1$ Hz is the frequency resolution of the ANPS. We thus find that the average noise power across all detectors is reduced by 74%. This algorithm also helps to mitigate the effects of the sea waves by removing roughly half of the noise power at sub-Hz frequencies. Consistently denoising the CUORE data using a combination of seismometers, accelerometers, and microphones could therefore help to offset the seasonally varying effect of the sea waves on the sensitivity to ($0\nu\beta\beta$) decay.

Finally, it is important to highlight that denoising has the greatest impact on the low-energy region of the spectrum. As in many other experiments, CUORE models the energy resolution of each detector as:

$$\sigma(E) = \sqrt{\sigma^2(0) + f(E)} \tag{13}$$

where $f(E)$ is a monotonically increasing function of energy with $f(0) = 0$[25]. The specific form of $f$ can vary between experiments and even between analyses. The denoising algorithm improves only the baseline resolution, $\sigma(0)$, so its effect is most significant at low energies, where $f(E)$ contributes the least. As a result, denoising is expected to provide the most benefit for low-energy analyses in CUORE, such as searches for dark matter interactions and other rare processes[32]. However, even in high-energy searches such as those for $0\nu\beta\beta$ decay, denoising provides an advantage by enabling lower energy thresholds, which in turn improves the ability to reject events in coincidence with low-energy signals, resulting in lower backgrounds.

## Data availability

The data used in this study include raw bolometer waveform data as well as higher-level processed data products derived from these waveforms for analysis, together with associated environmental and auxiliary data. These data are stored on internal computing clusters of the CUORE Collaboration and are not publicly available due to their size and collaboration access policies. No accession codes or persistent identifiers are associated with these datasets. The data are available for scientific purposes from the corresponding author upon reasonable request, who is responsible for handling data access inquiries.

## Acknowledgements

The CUORE Collaboration thanks the directors and staff of the Laboratori Nazionali del Gran Sasso and the technical staff of our laboratories. This work was supported by the Istituto Nazionale di Fisica Nucleare (INFN); the National Science Foundation under Grant Nos. NSF-PHY-0605119, NSF-PHY-0500337, NSF-PHY-0855314, NSF-PHY-0902171, NSF-PHY-0969852, NSF-PHY-1307204, NSF-PHY-1314881, NSF-PHY-1401832, NSF-PHY-1913374, and NSF-PHY-2412377; Yale University, Johns Hopkins University, and University of Pittsburgh. This material is also based upon work supported by the US Department of Energy (DOE) Office of Science under Contract Nos. DE-AC02-05CH11231, and DE-AC52-07NA27344; by the DOE Office of Science, Office of Nuclear Physics under Contract Nos. DE-FG02-08ER41551, DE-FG03-00ER41138, DE-SC0012654, DE-SC0020423, DE-SC0019316 and DE-SC0011091. This research used resources of the National Energy Research Scientific Computing Center (NERSC). This work makes use of both the DIANA data analysis and APOLLO data acquisition software packages, which were developed by the CUORICINO, CUORE, LUCIFER, and CUPID-0 Collaborations. The authors acknowledge the Advanced Research Computing at Virginia Tech and the Yale Center for Research Computing for providing computational resources and technical support that have contributed to the results reported within this paper. This study has been conducted using E.U. Copernicus Marine Service information and data from the INGV GIGS seismic station.

## Author contributions

All listed authors have contributed to the present publication. The different contributions span from the design and construction of the detector and the cryogenic system to the acquisition and analysis of data. The manuscript underwent an internal review process extended to the whole collaboration, and all authors approved its final version; the authors' names are listed alphabetically.

## Competing interests

The authors declare no competing interests.

## Additional information

**Correspondence** and requests for materials should be addressed to C. Bucci.

**Peer review information** : *Communications Physics* thanks Jeanne Wilson, Daniel Cookman and the other, anonymous, reviewer(s) for their contribution to the peer review of this work. A peer review file is available.

**D.Q. Adams[1], C. Alduino[1], K. Alfonso[2], A. Armatol[3], F. T. Avignone III[1], O. Azzolini[4], G. Bari[5], F. Bellini[6,7], G. Benato[8,9], M. Beretta[10,11], M. Biassoni[11], A. Branca[10,11], C. Brofferio[10,11], C. Bucci[9] ✉, J. Camilleri[2], A. Caminata[12], A. Campani[12,13], J. Cao[14], C. Capelli[3], S. Capelli[10,11], L. Cappelli[9], L. Cardani[7], P. Carniti[10,11], N. Casali[7], E. Celi[8,9], D. Chiesa[10,11], M. Clemenza[11], S. Copello[15], O. Cremonesi[11], R. J. Creswick[1], A. D'Addabbo[9], I. Dafinei[7], S. Dell'Oro[10,11], S. Di Domizio[12,13], S. Di Lorenzo[9], D. Q. Fang[14], M. Faverzani[10,11], E. Ferri[11], F. Ferroni[7,8], E. Fiorini[10,11,37], M. A. Franceschi[16], S. J. Freedman[3,17,38], S. H. Fu[9,14], B. K. Fujikawa[3], S. Ghislandi[8,9], A. Giachero[10,11], M. Girola[10,11], L. Gironi[10,11], A. Giuliani[18], P. Gorla[9] C. Gotti[11], P. V. Guillaumon[9,36], T. D. Gutierrez[19], K. Han[20], E. V. Hansen[17], K. M. Heeger[21], D. L. Helis[9], H. Z. Huang[22], M. T. Hurst[23], G. Keppel[4], Yu. G. Kolomensky[3,17], R. Kowalski[24], R. Liu[21], L. Ma[14,22], Y. G. Ma[14], L. Marini[9], R. H. Maruyama[21], D. Mayer[3,17,25], Y. Mei[3], M. N. Moore[21], T. Napolitano[16], M. Nastasi[10,11], C. Nones[26], E. B. Norman[27], A. Nucciotti[10,11], I. Nutini[11], T. O'Donnell[2], M. Olmi[9], B. T. Oregui[24], S. Pagan[21], C. E. Pagliarone[9,28], L. Pagnanini[8,9], M. Pallavicini[12,13], L. Pattavina[10,11], M. Pavan[10,11], G. Pessina[11], V. Pettinacci[7], C. Pira[4], S. Pirro[9], E. G. Pottebaum[21], S. Pozzi[11], E. Previtali[10,11], A. Puiu[9],**

**S. Quitadamo[8,9], A. Ressa[7], C. Rosenfeld[1], B. Schmidt[26], R. Serino[10,18], A. Shaikina[8,9], V. Sharma[23], V. Singh[17], M. Sisti[11], D. Speller[24], P. T. Surukuchi[23], L. Taffarello[29], C. Tomei[7], A. Torres[2], J. A. Torres[21], K. J. Vetter[3,17,25], M. Vignati[6,7], S. L. Wagaarachchi[3,17], B. Welliver[3,18], J. Wilson[1], K. Wilson[1], L. A. Winslow[25], F. Xie[14], T. Zhu[18], S. Zimmermann[30], S. Zucchelli[5,31], CUORE Collaboration*, L. Aragão[32], A. Armigliato[31], R. Brancaccio[31,33], F. del Corso[5,34], S. Castellaro[31], G. De Luca[35], S. di Sabatino[31], P. Ruggieri[31] & M. Zavatarelli[31]**

[1]Department of Physics and Astronomy, University of South Carolina, Columbia, SC, USA. [2]Center for Neutrino Physics, Virginia Polytechnic Institute and State University, Blacksburg, VA, USA. [3]Nuclear Science Division, Lawrence Berkeley National Laboratory, Berkeley, CA, USA. [4]INFN—Laboratori Nazionali di Legnaro Legnaro, Padova, Italy. [5]INFN—Sezione di Bologna, Bologna, Italy. [6]Dipartimento di Fisica, Sapienza Università di Roma, Roma, Italy. [7]INFN—Sezione di Roma, Roma, Italy. [8]Gran Sasso Science Institute, L'Aquila, Italy. [9]INFN—Laboratori Nazionali del Gran Sasso, Assergi (L'Aquila), Italy. [10]Dipartimento di Fisica, Università di Milano-Bicocca, Milano, Italy. [11]INFN—Sezione di Milano Bicocca, Milano, Italy. [12]INFN—Sezione di Genova, Genova, Italy. [13]Dipartimento di Fisica, Università di Genova, Genova, Italy. [14]Key Laboratory of Nuclear Physics and Ion-beam Application (MOE), Institute of Modern Physics, Fudan University, Shanghai, China [15]INFN—Sezione di Pavia, Pavia, Italy. [16]INFN—Laboratori Nazionali di Frascati, Frascati (Roma), Italy. [17]Department of Physics, University of California, Berkeley, CA, USA. [18]Université Paris-Saclay, CNRS/IN2P3, IJCLab, Orsay, France. [19]Physics Department, California Polytechnic State University, San Luis Obispo, CA, USA. [20]INPAC and School of Physics and Astronomy, Shanghai Jiao Tong University; Shanghai Laboratory for Particle Physics and Cosmology, Shanghai, China. [21]Wright Laboratory, Department of Physics, Yale University, New Haven, CT, USA. [22]Department of Physics and Astronomy, University of California, Los Angeles, CA, USA. [23]Department of Physics and Astronomy, University of Pittsburgh, Pittsburgh, PA, USA. [24]Department of Physics and Astronomy, The Johns Hopkins University, Baltimore, MD, USA. [25]Massachusetts Institute of Technology, Cambridge, MA, USA. [26]IRFU, CEA, Université Paris-Saclay, Gif-sur-Yvette, France. [27]Department of Nuclear Engineering, University of California, Berkeley, CA, USA. [28]Dipartimento di Ingegneria Civile e Meccanica, Università degli Studi di Cassino e del Lazio Meridionale, Cassino, Italy. [29]INFN—Sezione di Padova, Padova, Italy. [30]Engineering Division, Lawrence Berkeley National Laboratory, Berkeley, CA, USA [31]Dipartimento di Fisica e Astronomia, Alma Mater Studiorum—Università di Bologna, Bologna, Italy. [32]CMCC Foundation—Euro-Mediterranean Center on Climate Change, Bologna, Italy. [33]Department of Physics and Earth Science, University of Ferrara, Ferrara, Italy. [34]Istituto Nazionale di Fisica Nucleare—Sezione di Perugia, Perugia, Italy. [35]Istituto Nazionale di Geofisica e Vulcanologia, Osservatorio Nazionale Terremoti—Sede di L'Aquila, L'Aquila, Italy. [36]Present address: Instituto de Física, Universidade de São Paulo, São Paulo, Brazil. [38]Deceased: E. Fiorini. [38]Deceased: S. J. Freedman. *A list of authors and their affiliations appears at the end of the paper. ✉e-mail: cuore-spokesperson@lngs.infn.it

## CUORE Collaboration

**D. Q. Adams[1], C. Alduino[1], K. Alfonso[2], A. Armatol[3], F. T. Avignone III[1], O. Azzolini[4], G. Bari[5], F. Bellini[6,7], G. Benato[8,9], M. Beretta[10,11], M. Biassoni[11], A. Branca[10,11], C. Brofferio[10,11], C. Bucci[9] ✉, J. Camilleri[2], A. Caminata[12], A. Campani[12,13], J. Cao[14], C. Capelli[3], S. Capelli[10,11], L. Cappelli[9], L. Cardani[7], P. Carniti[10,11], N. Casali[7], E. Celi[8,9], D. Chiesa[10,11], M. Clemenza[11], S. Copello[15], O. Cremonesi[11], R. J. Creswick[1], A. D'Addabbo[9], I. Dafinei[7], S. Dell'Oro[10,11], S. Di Domizio[12,13], S. Di Lorenzo[9], D. Q. Fang[14], M. Faverzani[10,11], E. Ferri[11], F. Ferroni[7,8], E. Fiorini[10,11,37], M. A. Franceschi[16], S. J. Freedman[3,17,38], S. H. Fu[9,14], B. K. Fujikawa[3], S. Ghislandi[8,9], A. Giachero[10,11], M. Girola[10,11], L. Gironi[10,11], A. Giuliani[18], P. Gorla[9], C. Gotti[11], P. V. Guillaumon[9], T. D. Gutierrez[19], K. Han[20], E. V. Hansen[17], K. M. Heeger[21], D. L. Helis[9], H. Z. Huang[22], M. T. Hurst[23], G. Keppel[4], Yu. G. Kolomensky[3,17], R. Kowalski[24], R. Liu[21], L. Ma[14,22], Y. G. Ma[14], L. Marini[9], R. H. Maruyama[21], D. Mayer[3,17,25], Y. Mei[3], M. N. Moore[21], T. Napolitano[16], M. Nastasi[10,11], C. Nones[26], E. B. Norman[27], A. Nucciotti[10,11], I. Nutini[11], T. O'Donnell[2], M. Olmi[9], B. T. Oregui[24], S. Pagan[21], C. E. Pagliarone[9,28], L. Pagnanini[8,9], M. Pallavicini[12,13], L. Pattavina[10,11], M. Pavan[10,11], G. Pessina[11], V. Pettinacci[7], C. Pira[4], S. Pirro[9], E. G. Pottebaum[21], S. Pozzi[11], E. Previtali[10,11], A. Puiu[9], S. Quitadamo[8,9], A. Ressa[7], C. Rosenfeld[1], B. Schmidt[26], R. Serino[10,18], A. Shaikina[8,9], V. Sharma[23], V. Singh[17], M. Sisti[11], D. Speller[24], P. T. Surukuchi[23], L. Taffarello[29], C. Tomei[7], A. Torres[2], J. A. Torres[21], K. J. Vetter[3,17,25], M. Vignati[6,7], S. L. Wagaarachchi[3,17], B. Welliver[3,17], J. Wilson[1], K. Wilson[1], L. A. Winslow[25], F. Xie[14], T. Zhu[17], S. Zimmermann[30] & S. Zucchelli[5,31]**